# Modeling and Instability of Average Current Control

Chung-Chieh Fang

Taiwan Semiconductor Manufacturing Co., Hsinchu 300, Taiwan

Tel: +886-3-5781688 ext 1261, Fax: +886-3-5637386, Email: ccfangb@tsmc.com.tw

## Keywords

Modelling, DC power supplies

## Abstract

Dynamics and stability of average current control of DC-DC converters are analyzed by sampled-data modeling. Orbital stability is studied and it is found unrelated to the ripple size of the orbit. Compared with the averaged modeling, the sampled-data modeling is more accurate and systematic. An unstable range of compensator pole is found by simulations, and is predicted by sampled-data modeling and harmonic balance modeling.

## 1 Introduction

In this paper, dynamics and stability of average current control (ACC) of DC-DC converters are analyzed by sampled-data modeling and harmonic balance modeling. Previous works [1, 2, 3, 4, 5, 6] on ACC DC-DC converters are generally based on averaged modeling. In the averaged modeling, each circuit module is modeled *separately* and *approximately*. For example, the switches are approximated as a three-terminal model. The current loop is obtained using sampled-data dynamics and then approximated by continuous-time dynamics. The PWM modulator is approximated as a constant gain. The duty cycle, a *discrete-time* variable, is treated as a continuous-time variable. Then all of these approximate modules form a continuous-time model. Therefore, the *orbital* nature of the nominal periodic solution is lost. Instead, an equilibrium is obtained as the nominal solution, and graphical ripple analysis is used to determine stability.

In contrast, no such approximations are involved in the sampled-data modeling. The orbital nature of the nominal periodic solution is preserved, and exact orbital stability can be determined. Also, the sampled-data modeling focuses on *system* operations, especially *switching* actions. The switching action is very important for the derivation of system dynamics, and it is accurately formulated in the sampled-data modeling. Therefore, the advantage of the sampled-data modeling is that it is more *accurate* and *systematic*. Although sampled-data analysis of converters has been a topic of investigation for the past two decades [7, 8, 9, 10, 11], this powerful tool is not widely used. This paper has an aim to increase the appreciation and use of the sampled-data modeling.

Harmonic balance is another approach to analyze subharmonic oscillation (period-doubling bifurcation) in DC-DC converters [12]. By using this approach, a simple analytical form for stability is obtained and it implies existence of an unstable range of compensator pole, which is confirmed by simulations.

The remainder of the paper is organized as follows. In Section 2, the operation of average current control is reviewed and an instability example is given. In Section 3, a general sampled-data model for DC-DC converters developed in [13] is reviewed and applied to the ACC DC-DC converter. In Section 4, harmonic balance analysis developed in [12] is applied to study period-doubling bifurcation in the ACC buck converter. Conclusions are collected in Section 5.

## 2  Operation and Instability of Average Current Control

An ACC DC-DC converter is shown in Fig. 1. Let switching frequency $f_s = 1/T$, and $\omega_s = 2\pi f_s$. The ACC operation is as follows [3]: The inductor current $i_L$ is sensed by a resistor $R_s$ and compared with a current reference $v_r$ from a voltage loop (not shown). The difference is amplified by a compensator. A typical compensator is

$$H_c(s) = \frac{H_n(s)}{H_d(s)} = \frac{K_c(1 + \frac{s}{\omega_z})}{s(1 + \frac{s}{\omega_p})} \qquad (1)$$

The compensator output $y(t)$ is compared to a $T$-periodic ramp signal $h(t)$, which has $h(0) = V_l$ and $h(T^-) = V_h$, to generate a switching signal to the power stage. The nominal *steady-state* solution of a DC-DC converter is a $T$-*periodic orbit*, not an equilibrium point as depicted in the averaged modeling. Denote the orbit by $x^0(t)$, and the corresponding steady-state compensator output by $y^0(t)$. An orbit (like $x^0(t)$ or $y^0(t)$) is (orbitally) stable if a state trajectory stays in the orbit if it is perturbed. If the state trajectory eventually leaves the orbit, then the orbit is unstable. Previous works [1, 2, 4, 5] imply that $\omega_p$ and ripple size of $y$ affect stability linearly. From the following simulations, however, an unstable range of $\omega_p$ is found and the ripple size is *unrelated* to *orbital* stability.

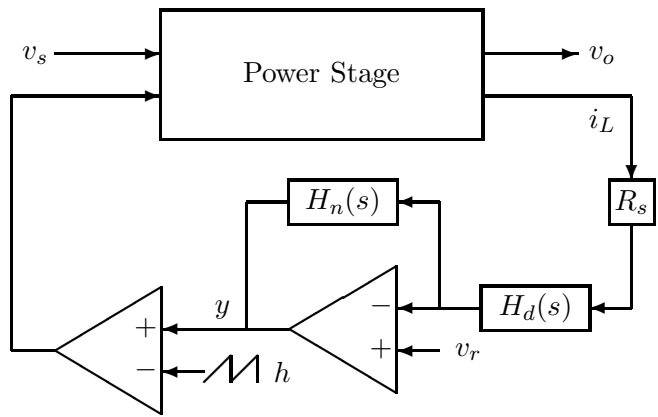

Figure 1: System diagram of an ACC DC-DC converter

**Example 1** Consider an ACC buck converter [2, p.114]. The system parameters are $v_s = 14$ V, nominal (set-point) output voltage $V_{\text{SET}} = 5$ V, $v_r = 0.5$, $f_s = 50$ kHz, $L = 46.1$ $\mu$H, $C = 380$ $\mu$F (with ESR $R_c = 0.02$ $\Omega$), $R = 1$ $\Omega$, $R_s = 0.1$ $\Omega$, $V_l = 0$, $V_h = 1$, $K_c = 75506$ and $\omega_z = 5652.9$ rad/s.

Compensator pole $\omega_p$ is varied from $0.14\omega_s$ to $0.81\omega_s$. The compensator output ripple size increases as $\omega_p$ increases. For $\omega_p < 0.19\omega_s$, $x^0(t)$ or $y^0(t)$ is stable. For $0.19\omega_s \leq \omega_p \leq 0.49\omega_s$, $x^0(t)$ is *unstable*. Take $\omega_p = 0.21\omega_s$ for example, $y^0(t)$ is unstable with small ripple (Fig. 2). For $\omega_p = 0.49\omega_s$, period-doubling bifurcation occurs, and the unstable $T$-periodic orbit (Fig. 3) and the stable $2T$-periodic orbit (Fig. 4) *coexits*. The two orbits have the *same* averaged duty cycle 0.357 and share the same *averaged* orbits. Based on the averaged model, they should have the same stability. However, their stabilities are different. For $\omega_p > 0.49\omega_s$, $x^0(t)$ or $y^0(t)$ is stable again. Take $\omega_p = 0.81\omega_s$ for example, $y^0(t)$ is stable with large ripple (Fig. 5). □

The orbital stability and the unstable range of $\omega_p$ will be explained by the sampled-data modeling and harmonic balance modeling discussed next.

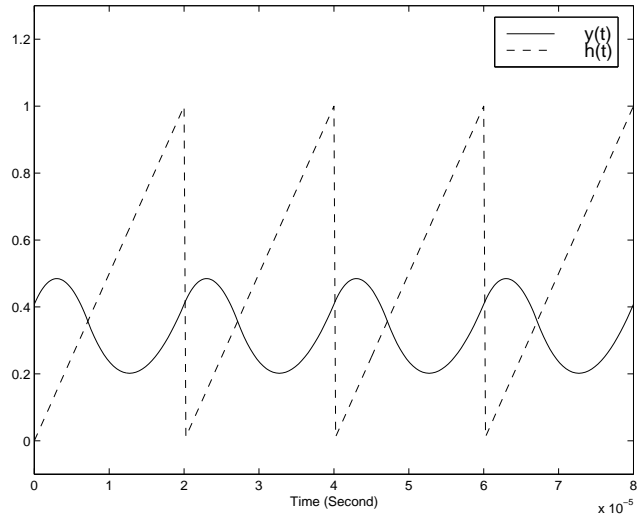

Figure 2: *Unstable* $y^0(t)$ with *small* ripple, $\omega_p = 0.21\omega_s$

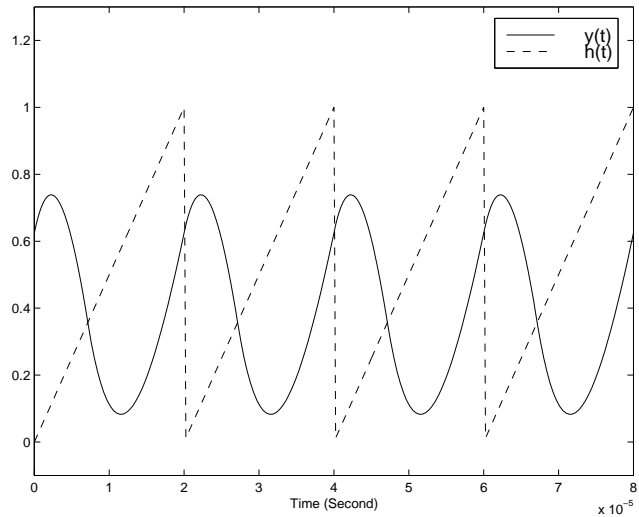

Figure 3: Unstable $y^0(t)$, $\omega_p = 0.49\omega_s$

## 3 Sampled-Data Analysis

The operation of average current control can be described *exactly* by a block diagram model developed in [13], which studied *closed-loop* DC-DC converters. Here the model is applied to an ACC DC-DC converter, with the voltage loop *open* and the reference signal $v_r$ considered as a control variable.

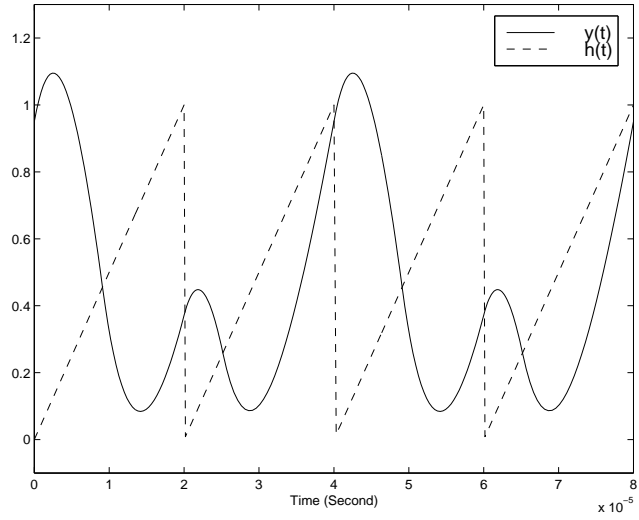

Figure 4: Stable $2T$-periodic orbit, $\omega_p = 0.49\omega_s$

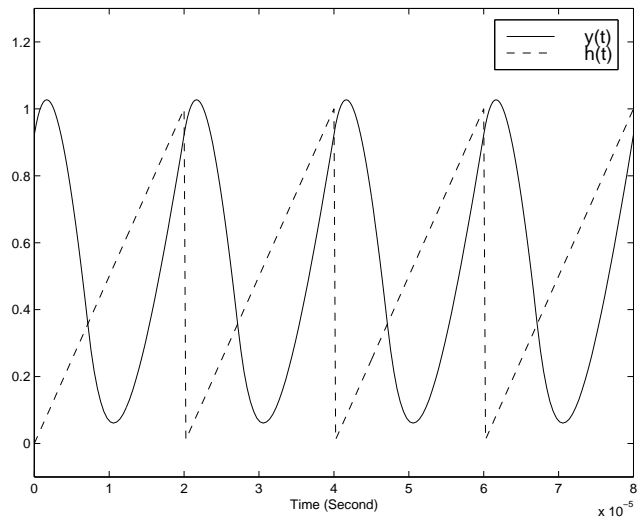

Figure 5: *Stable* $y^0(t)$ with *large* ripple, $\omega_p = 0.81\omega_s$

The block diagram model is shown in Fig. 6, and it can model many DC-DC converters, such as buck, boost, and Cuk converters, etc. In the model, $A_1, A_2 \in \mathbf{R}^{N \times N}$, $B_1, B_2 \in \mathbf{R}^{N \times 2}$, $C, E_1, E_2 \in \mathbf{R}^{1 \times N}$, and $D \in \mathbf{R}^{1 \times 2}$ are constant matrices, where $N$ is the state dimension, typically given by the number of energy storage elements in *both* of the power stage and the compensator. For example, $N = 3$ for a typical second order buck converter with a first order compensator. Within a clock period, the dynamics is switched between two stages, $S_1$ and $S_2$. Switching occurs when the ramp signal $h(t)$ intersects with the compensator output $y := Cx + Du \in \mathbf{R}$. The system is in $S_1$ when $y(t) \geq h(t)$, and switches to $S_2$ at instants when $y(t) < h(t)$. Here the switching operation is modeled *exactly*, without *seperate* and *graphical* approximations as in the averaged model.

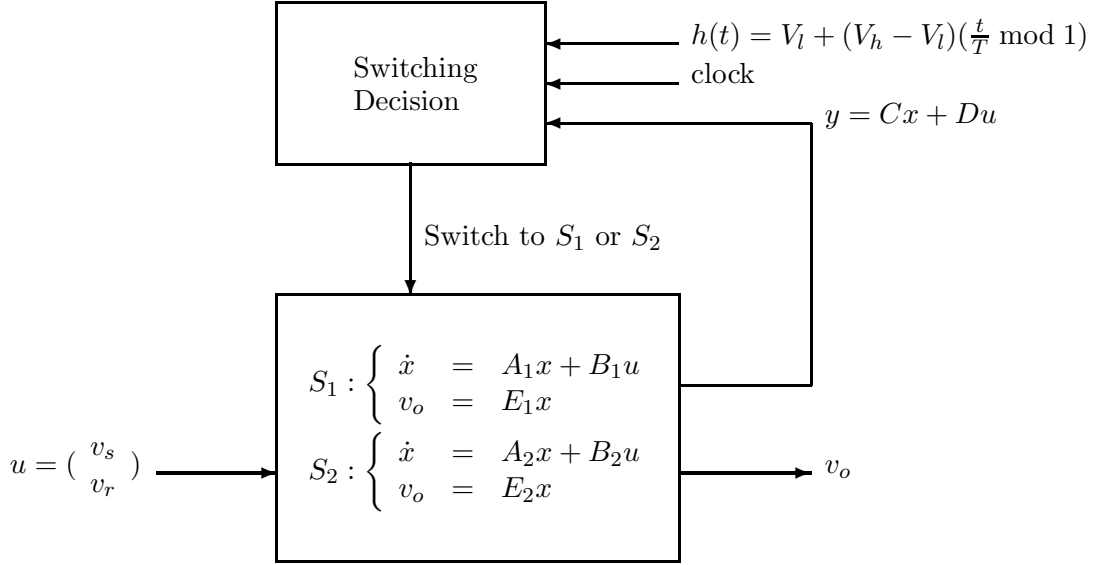

Figure 6: Block diagram model for ACC DC-DC converter

**Example 2** Consider Example 1. Let the state $x = (i_L, v_C, v_{e1}, v_{e2})'$, where $i_L$ is the inductor current; $v_C$ is the capacitor voltage; $v_{e1}$ and $v_{e2}$ are the states of the compensator. Then,

$$A_1 = A_2 = \begin{bmatrix} \frac{-RR_c}{(R+R_c)L} & \frac{-R}{(R+R_c)L} & 0 & 0 \\ \frac{R}{(R+R_c)C} & \frac{-1}{(R+R_c)C} & 0 & 0 \\ 0 & 0 & 0 & 1 \\ -\omega_p R_s & 0 & 0 & -\omega_p \end{bmatrix}$$

$$B_1 = \begin{bmatrix} \frac{1}{L} & 0 \\ 0 & 0 \\ 0 & 0 \\ 0 & \omega_p \end{bmatrix} \qquad B_2 = \begin{bmatrix} 0 & 0 \\ 0 & 0 \\ 0 & 0 \\ 0 & \omega_p \end{bmatrix}$$

$$C = \begin{bmatrix} 0 & 0 & K_c & \frac{K_c}{\omega_z} \end{bmatrix} \qquad D = \begin{bmatrix} 0 & 1 \end{bmatrix}$$

$$E_1 = E_2 = \begin{bmatrix} \frac{RR_c}{R+R_c} & \frac{R}{R+R_c} & 0 & 0 \end{bmatrix}$$

□

Let $x_n = x(nT)$. Assume the switching frequency is high enough that $u = (v_s, v_r)' \in \mathbf{R}^{2 \times 1}$ can be considered constant *within* the cycle, and denote its value by $u_n = (v_{sn}, v_{rn})'$, which can be varied from cycle to cycle. The notation $v_{sn}$, instead of $v_{s,n}$, is used for brevity.

The periodic orbit $x^0(t)$ in Fig. 6 corresponds to a fixed point $x^0(0)$ in the sampled-data dynamics. Let the steady-state duty cycle be $D_c$ and $d = D_c T$. A typical periodic solution $x^0(t)$ is shown in Fig. 7, where $\dot{x}^0(d^-) = A_1 x^0(d) + B_1 u$ and $\dot{x}^0(d^+) = A_2 x^0(d) + B_2 u$ denote the time derivative of $x^0(t)$ at $t = d^-$ and $d^+$, respectively.

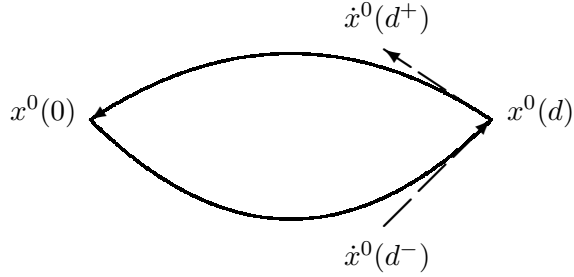

Figure 7: A typical periodic solution $x^0(t)$ of a DC-DC converter in state space

Using a hat ˆ to denote small perturbations (e.g., $\hat{x}_n = x_n - x^0(0)$). From [13], the linearized sampled-data dynamics is

$$\hat{x}_{n+1} = \Phi \hat{x}_n + \Gamma \hat{u}_n = \Phi \hat{x}_n + \Gamma_1 \hat{v}_{sn} + \Gamma_2 \hat{v}_{rn} \qquad (2)$$

where $\Phi \in \mathbf{R}^{N \times N}$ and $\Gamma = [\Gamma_1, \Gamma_2] \in \mathbf{R}^{N \times 2}$ are

$$\begin{aligned}
\Phi &= e^{A_2(T-d)}(I - \frac{((A_1 - A_2)x^0(d) + (B_1 - B_2)u)C}{C(A_1 x^0(d) + B_1 u) - \dot{h}(d)}) e^{A_1 d} \\
&= e^{A_2(T-d)}(I - \frac{(\dot{x}^0(d^-) - \dot{x}^0(d^+))C}{C\dot{x}^0(d^-) - \dot{h}(d)}) e^{A_1 d} \qquad (3)\\
\Gamma &= e^{A_2(T-d)}(\int_0^d e^{A_1 \sigma} d\sigma B_1 - \frac{\dot{x}^0(d^-) - \dot{x}^0(d^+)}{C\dot{x}^0(d^-) - \dot{h}(d)}(C \int_0^d e^{A_1 \sigma} d\sigma B_1 + D)) + \int_0^{T-d} e^{A_2 \sigma} d\sigma B_2
\end{aligned}$$

Local *orbital* stability of the converter is determined by the eigenvalues of $\Phi$. The periodic solution $x^0(t)$ is asymptotically orbitally stable if all of the eigenvalues of $\Phi$ are inside the unit circle of the complex plane.

**Example 3** Consider again Example 1. The eigenvalues of $\Phi$ are calculated as $\omega_p$ is varied from $0.14\omega_s$ to $0.81\omega_s$. For $\omega_p = 0.14\omega_s$, all eigenvalues are inside the unit circle and $x^0(t)$ is stable. As $\omega_p$ is increased, the eigenvalue trajectory leaves the unit circle through -1 when $\omega_p = 0.19\omega_s$, and enters again the unit circle when $\omega_p = 0.49\omega_s$. This explains exactly the unstable range of $\omega_p$ reported in Example 1. □

**Example 4** For comparison, the linearized dynamics of state-space *averaged* model of Example 1 is derived:

$$\dot{\hat{x}} = (A_1 + \frac{(B_1 - B_2)uC}{V_h - V_l})\hat{x} \qquad (4)$$

For $\omega_p$ varied from $0.14\omega_s$ to $0.81\omega_s$, the poles of (4) are in the left half of complex plane and the converter should be stable according to the average model. This contradicts the simulation results in Example 1. □

The dynamics (2) can be also used to derive some useful transfer functions. The reference signal $v_r$ is used as a control variable. Let $E = (E_1 + E_2)/2$. From Eq. (2), the control-to-output

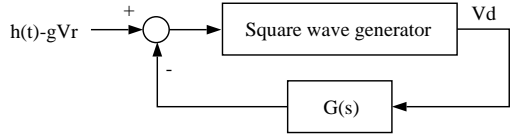

Figure 8: Unified dynamic model of buck converter

voltage transfer function is

$$T_{oc}(z) = \frac{\hat{v}_o(z)}{\hat{v}_r(z)} = E(zI - \Phi)^{-1}\Gamma_2 \quad (5)$$

Given a transfer function in $z$ domain, say $T(z)$, its effective frequency response is $T(e^{j\omega T})$, which is valid in the frequency range $|\omega| < \frac{\pi}{T}$.

Also from Eq. (2), audio-susceptibility is

$$T_{os}(z) = \frac{\hat{v}_o(z)}{\hat{v}_s(z)} = E(zI - \Phi)^{-1}\Gamma_1 \quad (6)$$

Other transfer functions, such as control-to-inductor current transfer function and output impedance can be derived similarly.

## 4 Harmonic Balance Analysis of Period-Doubling Bifurcation

The unstable range of $\omega_p$ can also be predicted by harmonic balance modeling. In [12], harmonic balance modeling is applied to study period-doubling bifurcation of a buck converter under *voltage* or *current* mode control. The buck converter model is shown in Fig. 8. Here, the same model can be applied to study an $ACC$ buck converter, with $G(s) = R_s H_c(s) G_1(s)$, where $G_1(s)$ is duty-cycle-to-inductor-current transfer function

$$\frac{RCs + 1}{RLCs^2 + Ls + R} \quad (7)$$

Let the critical source voltage be $V_s^*$ such that period-doubling bifurcation occurs for $v_s > V_s^*$. A larger $V_s^*$ implies a larger stable operating range of source voltage. According to [12], a good estimate of $V_s^*$ is

$$V_s^* \approx \frac{V_h - V_l}{2\mathrm{Re}[G(j\omega_s) - G(\frac{j\omega_s}{2})]} \quad (8)$$

For an ACC buck converter with $\omega_s \gg 1/\sqrt{LC}$ and $1/RC$, Eq. (8) can be simplified as

$$V_s^* \approx \frac{2(V_h - V_l)L\omega_z\omega_s}{3R_s K_c}\phi(k) \quad (9)$$

where $k = \omega_p/\omega_s$ and

$$\phi(k) = \frac{(1 + k^2)(0.25 + k^2)}{k} \quad (10)$$

The function $\phi(k) = \phi(\omega_p/\omega_s)$ is shown in Fig. 9 and has a minimum at $\omega_p = 0.38\omega_s$. From (9), the following operating conditions are desired to have a larger stable range of source voltage: larger $\phi$, ramp amplitude $(V_h - V_l)$, $L$, $\omega_z$, $\omega_s$; or smaller gains $R_s$ and $K_c$. This gives insight on how these parameters affect (period-doubling) instability. A guideline of previous works [5] on ACC is $\omega_p = 0.5\omega_s$. From Fig. 9, this choice of $\omega_p$ results in smaller $\phi$ and smaller stable operating range of source voltage. Setting $\omega_p = \omega_s$ instead (as an example) will have two times larger operating range. Let

$$V_s^{min} := V_s^*|_{k=0.38} \approx \frac{0.79(V_h - V_l)L\omega_z\omega_s}{R_s K_c} \qquad (11)$$

In (9), $V_s^*$ is a function of $\omega_p$. If there exists a $\omega_p$ such that $V_s^* \leq v_s$, then period-doubling occurs. Therefore, a larger $V_s^{min}$ is desired and the relationship between $V_s^{min}$ and $v_s$ determines whether there exists an unstable range of $\omega_p$:

**Theorem 1** *Consider an ACC converter with a source voltage $v_s$. The voltage loop is open and the current loop has a compensator with a pole $\omega_p$ and a zero. If $v_s \geq V_s^{min}$, then there exists an unstable range of $\omega_p$ which results in period-doubling instability. If $v_s < V_s^{min}$, then $\omega_p$ does not affect the stability.*

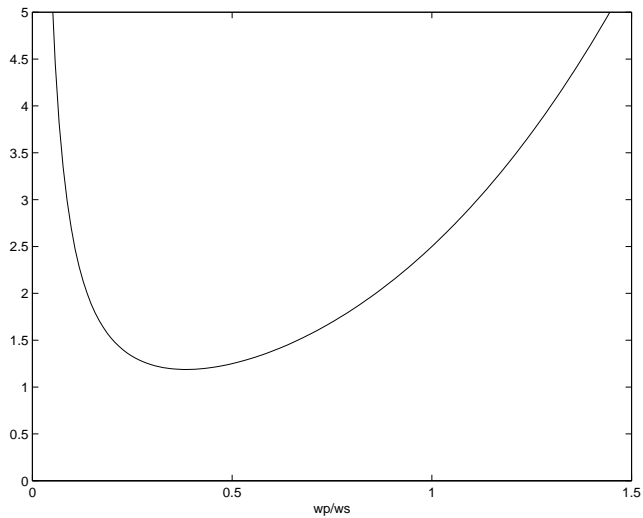

Figure 9: Function $\phi(k) = \phi(\omega_p/\omega_s)$

These two cases are illustrated in the following two examples.

**Example 5** In Example 1, $V_s^{min} = 8.57 < v_s$, therefore an unstable range of $\omega_p$ exists which results in period-doubling. This agrees with the simulation in Example 1. □

From Examples 3-5, both sampled-data and harmonic balance models, but not the averaged model, predict accurately about period-doubling.

**Example 6** Consider another ACC buck from [3, p.982]. The system parameters are $v_s = 5$ V, $V_{\text{SET}} = 2$ V, $v_r = 0.279$, $f_s = 180$ kHz, $L = 13$ $\mu$H, $C = 750$ $\mu$F (with ESR $R_c = 5$ m$\Omega$), $R = 0.43$ $\Omega$, $R_s = 0.06$ $\Omega$, $V_l = 0$, $V_h$=2.7, $K_c = 98000$, and $\omega_z = 6723$ rad/s.

Here $V_s^{min} = 35.86 > v_s$ and *period-doubling* does not occur for any value of $\omega_p$. However, other kind of instabilities may exist. Generally, $\omega_p$ is greater than $\omega_z$. To show rich dynamics in the converter, let $\omega_p = 5655 < \omega_z$. The nominal $y^0(t)$ is shown in Fig. 10. A pair of complex conjugate eigenvalues of the *sampled-data* model are outside the unit circle, which implies occurrence of a Neimark bifurcation [11]. Therefore, the orbit of $y^0(t)$ is *unstable* although it has very *small* ripple. Also, a pair of eigenvalues of the state-space *averaged* model (2) are calculated and they are found on the right half plane. Therefore, in this example, the averaged model (4) also predicts the Neimark bifurcation correctly. Generally, the Neimark bifurcation, but not the period-doubling bifurcation, can be predicted by the averaged model. □

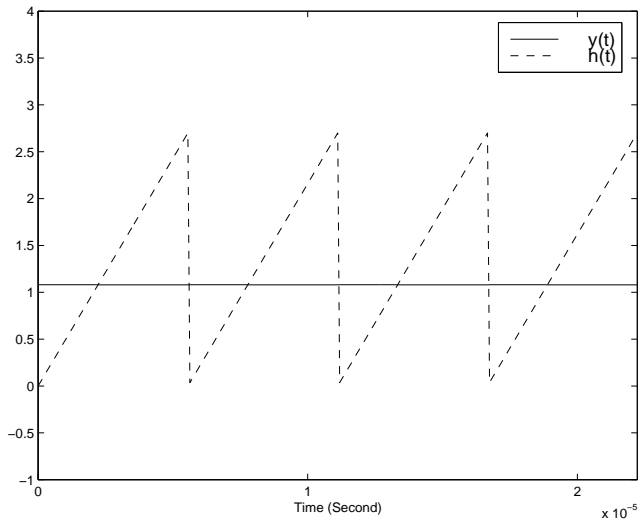

Figure 10: Unstable $y^0(t)$ with almost no ripple

## 5  Conclusion

Sampled-data modeling is applied to DC-DC converters under average current control. The *orbital* nature of the nominal periodic solution is preserved. Orbital stability is studied and is unrelated to the ripple size of the orbit. Various transfer functions are derived. In the sampled-data model, the system dynamics is derived *directly* and *exactly* from discrete switching operations. In the averaged model, however, each circuit module (e.g., the switch, PWM modulator and current sensing) is modeled *separately* and *approximately*. Therefore, the sampled-data modeling is more accurate and systematic than the averaged modeling.

An unstable range of compensator pole which results in period-doubling instability is found by simulations, which can be predicted by the sampled-data modeling and harmonic balance modeling. This nonlinear effect can not be predicted by the averaged modeling. How different system parameters affect stability is derived. For example, a critical source voltage $V_s^*$ is derived. The relationship of its minimum value $V_s^{min}$ with the source voltage determines whether an unstable range of compensator pole exists (Theorom 1).

Two typical instabilities in DC-DC converters are period-doubling bifurcation and Neimark bifurcation, and *both* can be accurately predicted by the sampled-data modeling. The period-doubling bifurcation can be also predicted by the harmonic balance modeling. The instability predicted by the averaged modeling is generally the Neimark bifurcation, but not the period-doubling bifurcation.